\documentclass[epj]{svjour}
\usepackage{graphics}
\usepackage{caption}
\usepackage{subfigure}
\usepackage{textcomp}
\usepackage{nicefrac}
\usepackage[switch]{lineno}

\begin{document}
\title{A new approach to monitor $^{13}$C-targets degradation {\em in situ} \\
for $^{13}$C($\alpha$,n)$^{16}$O cross-section measurements at LUNA}

\author{G.\,F. Ciani\inst{1,2,3} \and L. Csedreki\inst{1,2} \and J. Balibrea-Correa\inst{4,5} \and A. Best\inst{4,5} \and M. Aliotta\inst{6} \and F. Barile\inst{7} \and D. Bemmerer\inst{8} \and A.~Boeltzig\inst{1,2} \and C. Broggini\inst{9} \and C.\,G. Bruno\inst{6} \and A. Caciolli\inst{9,10} \and F. Cavanna\inst{11} \and T. Chillery\inst{6} \and P. Colombetti\inst{12,13} \and P.~Corvisiero\inst{11,14} \and T. Davinson\inst{6} \and R.~Depalo\inst{9} \and A. Di~Leva\inst{4,5} \and L. Di Paolo\inst{2} \and Z. Elekes\inst{3} \and F. Ferraro\inst{11,14} \and E.\,M. Fiore\inst{7,15} \and A. Formicola\inst{2} \and Zs. F\"ul\"op\inst{3} \and G. Gervino\inst{12,13} \and A. Guglielmetti\inst{16,17} \and C. Gustavino\inst{18} \and Gy. Gy\"urky\inst{3} \and G. Imbriani \inst{4,5} \and M.~Junker\inst{2} \and I. Kochanek \inst{2} \and M. Lugaro\inst{19} \and P.~Marigo\inst{9,10} \and E. Masha\inst{16,17} \and R. Menegazzo\inst{9} \and V. Mossa\inst{7} \and F.\,R.\ Pantaleo\inst{7,20} \and V.~Paticchio\inst{7} \and R. Perrino\inst{7,}\thanks{Permanent address: Istituto Nazionale di Fisica Nucleare, Sezione di Lecce, Via Arnesano, 73100 Lecce, Italy} \and D.~Piatti\inst{9,10}\and P. Prati\inst{11,14} \and L. Schiavulli\inst{7,15} \and K. St\"ockel\inst{8,21}
\and O. Straniero\inst{2,22} \and T. Sz\"ucs\inst{8} \and M.\,P.\ Tak\'acs\inst{8,21,}\thanks{Current address: Physikalisch-Technische Bundesanstalt, Bundesallee 100, 38116 Braunschweig, Germany} \and F.~Terrasi\inst{23} \and D. Trezzi\inst{16,17} \and S. Zavatarelli\inst{11}}
\authorrunning{Ciani et al.}
\titlerunning{A new approach to monitor $^{13}$C target degradation}
%
\offprints{G.F. Ciani, L. Csedreki} 
\mail{giovanni.ciani@lngs.infn.it \and laszlo.csedreki@lngs.infn.it}
\institute{Gran Sasso Science Institute, Viale F. Crispi 7, 67100 L'Aquila, Italy \and Istituto Nazionale di Fisica Nucleare Laboratori Nazionali del Gran Sasso (LNGS), Via G. Acitelli 22, 67100 Assergi, Italy \and Institute for Nuclear Research (Atomki), PO Box 51, 4001 Debrecen, Hungary \and Universit\`a degli Studi di Napoli ``Federico II'', Dipartimento di Fisica ``E. Pancini'', Via Cintia 21, 80126 Napoli, Italy \and Istituto Nazionale di Fisica Nucleare, Sezione di Napoli, Via Cintia 21, 80126 Napoli, Italy \and SUPA, School of Physics and Astronomy, University of Edinburgh, Peter Guthrie Tait Road, EH9 3FD Edinburgh, United Kingdom \and Istituto Nazionale di Fisica Nucleare, Sezione di Bari, Via E. Orabona 4, 70125 Bari, Italy  \and Helmholtz-Zentrum Dresden-Rossendorf, Bautzner Landstra\ss{}e 400, 01328 Dresden, Germany  \and Istituto Nazionale di Fisica Nucleare, Sezione di Padova, Via F. Marzolo 8, 35131 Padova, Italy \and Universit\`a degli Studi di Padova, Via F. Marzolo 8, 35131 Padova, Italy \and Istituto Nazionale di Fisica Nucleare, Sezione di Genova, Via Dodecaneso 33, 16146 Genova, Italy \and Universit\`a degli Studi di Torino, Via P. Giuria 1, 10125 Torino, Italy \and Istituto Nazionale di Fisica Nucleare, Sezione di Torino, Via P. Giuria 1, 10125 Torino, Italy \and Universit\`a degli Studi di Genova, Via Dodecaneso 33, 16146 Genova, Italy \and Universit\`a degli Studi di Bari, Dipartimento Interateneo di Fisica, Via G. Amendola 173, 70126 Bari, Italy \and Universit\`a degli Studi di Milano, Via G. Celoria 16, 20133 Milano, Italy \and Istituto Nazionale di Fisica Nucleare, Sezione di Milano, Via G. Celoria 16, 20133 Milano, Italy \and Istituto Nazionale di Fisica Nucleare, Sezione di Roma, Piazzale A. Moro 2, 00185 Roma, Italy \and Konkoly Observatory, Research Centre for Astronomy and Earth Sciences, Konkoly Thege Miklós út 15-17, H-1121 Budapest, Hungary\and Politecnico di Bari, Dipartimento Interateneo di Fisica, Via G. Amendola 173, 70126 Bari, Italy \and Technische Universit\"at Dresden, Institut f\"ur Kern- und Teilchenphysik, Zellescher Weg 19, 01069 Dresden, Germany \and INAF Osservatorio Astronomico d'Abruzzo, Via Mentore Maggini, 64100 Teramo, Italy \and Universit\`a degli Studi della Campania L. Vanvitelli, Dipartimento di Matematica e Fisica, Via Lincoln 5 - 81100 Caserta, Italy}

\date{Received: date / Revised version: date}
%
\abstract{
Direct measurements of reaction cross-sections at astrophysical energies often require the use of solid targets able to withstand high ion beam currents for extended periods of time. Thus, monitoring target thickness, isotopic composition, and target stoichiometry during data taking is critical to account for possible target modifications and to reduce uncertainties in the final cross-section results.
A common technique used for these purposes is the Nuclear Resonant Reaction Analysis (NRRA), which however requires that a narrow resonance be available inside the dynamic range of the accelerator used. In cases when this is not possible, as for example the $^{13}$C($\alpha$,n)$^{16}$O reaction recently studied at low energies at the Laboratory for Underground Nuclear Astrophysics (LUNA) in Italy, alternative approaches must be found.
Here, we present a new application of the shape analysis of primary $\gamma$ rays emitted by the $^{13}$C(p,$\gamma$)$^{14}$N radiative capture reaction. This approach was used to monitor $^{13}$C target degradation {\em in situ} during the $^{13}$C($\alpha$,n)$^{16}$O data taking campaign.  
The results obtained are in agreement with evaluations subsequently performed at Atomki (Hungary) using the  NRRA method.
A preliminary application for the extraction of the $^{13}$C($\alpha$,n)$^{16}$O reaction cross-section at one beam energy is also reported.
} 
\authorrunning{Ciani et. al}
\maketitle
\section{Introduction}
Knowledge of the stoichiometric composition of solid state targets and their behaviour during ion beam irradiation is of great importance in various fields of ion beam physics, from material analysis to nuclear astrophysics \cite{JESUS2001229,Cruz,SILVA2017135,Caciolli}.
The main goal of the latter is to measure nuclear reaction cross-sections at, or near, the energy region of astrophysical interest (the so-called Gamow window), typically of the order of hundreds of keV or less. Since cross-sections drop exponentially with decreasing energy in this energy region, counting rates can be of the order of one event per hour or lower. 
Therefore, high beam currents (hundreds of $\mu$A) and long irradiation times (weeks or months) are often necessary to achieve high enough signal-to-noise ratios for a successful cross-section measurement at low energies.
Yet, target modification processes (such as diffusion, melting, sputtering or contamination of target surface \cite{Paine1985,Radek2017}) that occur under intense beam irradiation may result in significant changes of target composition and/or stoichiometry as a function of irradiation depth \cite {Wang1984} and an {\em in-situ} monitoring of target properties is generally required. 

Typically, this is achieved by using the well-established Nuclear Resonant Reaction Analysis (NRRA) (see, for example, \cite{tesmer1995handbook,MCGLONE1991201} and refs. therein), which requires a narrow
resonance\footnote{A narrow resonance is defined as one whose total width $\Gamma$ is much smaller that the target thickness $\Delta E$ in energy units. The latter represents the energy lost by the ion beam in going through the target and depends on the initial beam energy as well as on the target composition and physical thickness.} to exist in the reaction of interest and to be accessible within the dynamic range of the particle accelerator. 
If no resonance is present or accessible, for example because of beam energy restrictions, other methods must be employed. 

This was the case of the astrophysically important $^{13}$C($\alpha$,n)$^{16}$O reaction \cite{Cristallo2018} recently studied in direct kinematics at the Laboratory for Underground Nuclear Astrophysics (LUNA) \cite{FORMICOLA,Review2018} of the Laboratori Nazionali del Gran Sasso (LNGS), INFN, Italy.
Because of the small cross-sections involved at the energies investigated ($E_\alpha = 305-400$~keV), intense $\alpha$-particle beams were needed, leading to severe target degradation and frequent target replacements.
Unfortunately, no resonances exist in the $^{13}$C($\alpha$,n)$^{16}$O reaction at $E_\alpha < 400$~keV and the NRRA method could not be used to monitor the state of $^{13}$C targets during irradiation.
Alternatively, one could use a proton beam, also available at LUNA,
on the same targets and exploit the $^{13}$C(p,$\gamma$)$^{14}$N reaction for NRRA analysis. However, also in this case no resonance exists that can be accessed with the 400~kV accelerator, hence a new approach to monitor the deterioration of $^{13}$C targets during $\alpha$-beam irradiation had to be used. 

In this paper, we report about an innovative application of the so-called $\gamma$-shape analysis \cite{Imbriani2005}. The approach consists in a detailed study of the shape of the $\gamma$-ray lines emitted in the radiative proton-capture process $^{13}$C(p,$\gamma$)$^{14}$N so as to periodically check both the thickness and stoichiometry of $^{13}$C targets used during the $^{13}$C($\alpha$,n)$^{16}$O campaign at LUNA. 
To validate the approach developed here, complementary NRRA measurements were also performed off-site (at Atomki in Debrecen, Hungary) on some targets, both before and after $\alpha$-beam irradiation at LUNA. 

The paper is organized as follows: 
first, we describe the NRRA technique used to characterize $^{13}$C targets at Atomki (sect.  \ref{section:targets-characterization}); then, we present the $\gamma$-shape approach applied to a primary transition in the $^{13}$C(p,$\gamma$)$^{14}$N reaction to assess target deterioration during the $^{13}$C($\alpha$,n)$^{16}$O campaign at LUNA (sect. \ref{sec:methodGamma}); 
and finally, we report the results of the validation procedure (sect. \ref{sec:validation}), together with a preliminary  application of the $\gamma$-shape analysis to the evaluation of the $^{13}$C($\alpha$,n)$^{16}$O reaction cross-section (sect. \ref{sec:cross-section}).

\section{Reaction yields and target properties: The NRRA approach} 
\label{section:targets-characterization}
The NRRA method is frequently used in measurements of reaction cross-sections of astrophysical interest and has already been extensively exploited in previous studies at LUNA \cite{Imbriani2005,Caciolli2011,Bruno2016,Bruno2019,Boeltzig2019}.

Briefly, the yield $Y$ of a nuclear reaction can be determined from experimental quantities as \cite{rolfs1988cauldrons}:
\begin{equation}
\label{eq:yieldexp}
    Y=\frac{N_{\rm R}}{N_{\rm b}}
\end{equation}
where $N_{\rm R}$ is the number of reactions (producing either particles or $\gamma$ rays) and $N_{\rm b}$ is the number of beam particles incident on the target.
The latter quantity can be determined as $Q/eq$, where $Q$ is the charge accumulated on target during beam irradiation, $e$ is the elementary charge and $q$ is the charge state of the projectile.
On the other hand, $Y$ is a function of the reaction cross-section $\sigma$ and the number $N_{\rm A}$ of active nuclei\footnote{For targets consisting of chemical compounds, active nuclei are defined as those of a given species that take part in the nuclear reaction under study. All other nuclear species present in the target do not contribute to the reaction yield and are regarded as inactive.} (per square centimetre) in the target. 

For targets of thickness $\Delta E$, corresponding to the energy lost by a beam of initial energy $E_0$ in traversing the target, 
and taking into account the energy dependence of the cross-section, the relationship between $Y$ and the cross-section $\sigma$ (at an energy $E$ within the target) can be expressed as \cite{rolfs1988cauldrons}:
\begin{equation}
    \label{eq:cross_sect}
    Y(E_{0}) = \int_{E_{0}-\Delta E}^{E_{0}}
    \frac{\sigma(E)}{\epsilon(E)}
    dE
\end{equation}

Here, $\epsilon$ is the so-called {\em stopping power} which, for a given beam ion and energy, depends only on the chemical composition and stoichiometry of the target. 
For compound targets containing both active and inactive nuclei, the effective stopping power $\epsilon_{\rm eff}$ is used instead, which can be parametrized using the Bragg's addition rule\footnote{For the present work a target composed of $^{13}$C and Ta was assumed (see sect. \ref{sec:targets-preparation}) and further corrections to Bragg's rule, typically required for carbon compounds with O and H, can safely be neglected.} \cite{Iliadis}:

\begin{equation}
    \label{eq:eff_stop}
    \epsilon_{\rm eff}(E)= \epsilon_{\rm A}(E) + 
    \sum_{\rm i}\frac{N_{\rm I_{i}}}{N_{\rm A}}\epsilon_{\rm I_i}(E)
\end{equation}
Here $\nicefrac{N_{\rm I}}{N_{\rm A}}$ is the ratio between inactive and active nuclei, and $\epsilon_{\rm A}$ and $\epsilon_{\rm I}$ are the stopping powers of the corresponding (active and inactive) pure materials. Their values are available in the literature and can be calculated using SRIM \cite{Ziegler2008}.

If the nuclear reaction cross-section is well known, a measurement of the yield (eq. \ref{eq:cross_sect}) can be used to experimentally determine the effective stopping power and thus to monitor the degree of deterioration of the target during beam irradiation. 
In particular, the NRRA method exploits the existence of a narrow and isolated resonance in a given reaction, whose cross-section is known and can be well described by the Breit-Wigner expression, $\sigma_{\rm BW}$ \cite{Iliadis}.
By measuring the yield as a function of beam energies in the proximity of the resonance and for targets of thickness $\Delta E$ much larger than the resonance width $\Gamma$, a characteristic resonance yield curve is obtained (see for example  Fig.\ref{fig:Yield_curve_NRRA_final}), which contains information about the target thickness and composition. 
Specifically, the height of the yield plateau depends on the target stoichiometry, while the FWHM of the yield profile provides a measure of the target thickness. If either or both the target thickness and stoichiometry change as a result of intense ion beam bombardment, so will the shape of the (thick-target, resonant) yield profile and repeated resonance scans can be used to quantify the degree of target deterioration.

\subsection{NRRA measurements at Atomki}
\label{sec:targets-preparation}
Solid targets were produced by evaporating 99\% enriched $^{13}$C powder (by Sigma Aldrich) on 4~cm diameter tantalum backings. In order to remove traces of light elements from the Ta surface, a cleaning procedure \cite{cleaning} with citric acid solution was used before evaporating the targets. 
The evaporation was performed by the electron gun technique using a Leybold UNIVEX 350 vacuum evaporator at Atomki.  
The vacuum chamber of the evaporator consists of a copper melting pot, an adjustable arm used to hold the tantalum disk at 10~cm from the melting pot, and an electron gun (similar to the setup described in \cite{WOLFE2000142}).
An oscillator quartz mounted inside the vacuum chamber at 15~cm from the melting pot was used to monitor the evaporation.

NRRA measurements were carried out at the 2~MV Medium-Current Plus Tandetron Accelerator \cite{Rajta2018} at Atomki immediately after target  production.
For these measurements, a narrow resonance in the $^{13}$C(p,$\gamma$)$^{14}$N reaction ($Q=7550.56$~keV) was used. The resonance is located at a proton beam energy
$E_{\rm p} = (1747.6 \pm 0.9)$~keV and has a width $\Gamma =$($135\pm8$)~eV \cite{Ajzenberg1991}. 
Thus, resonance scans were performed at beam energies in the range $E_{\rm p}=1742-1770$~keV.

Targets were irradiated with typical proton beam currents of $i=500 $~nA, covering a beam spot size of about 5~mm diameter. 
Given the low beam intensity on target, neither a cooling system nor a cold trap were needed for this setup.
The target chamber was isolated from other beam-line components and acted as a Faraday cup for charge integration. 
An electrically insulated collimator biased to $-300$~V was placed at the entrance of the chamber to suppress secondary electrons.
A 100\% relative efficiency n-type coaxial HPGe detector was mounted in close geometry, at a distance of about 3~cm from the target, and at 0$^{\circ}$ with respect to the beam axis.

Spectra of the emitted $\gamma$ rays were collected with an ORTEC MCA (model ASPEC 927) and the ORTEC MAESTRO software.
The region of interest (ROI) in the $\gamma$-ray spectra was set to $E_{\gamma}=8.0-9.4$~MeV ($E_\gamma \approx E_{\rm c.m.}+Q$) so as to include both the full-energy peak and the single- and double-escape peaks of the direct capture transition to the ground state of the $^{14}$N compound nucleus. 
Given the magnitude of the resonant cross-section ($\sigma_{\rm BW} \simeq 10$~mb \cite{TENDL2017}), it was possible to reach a statistical uncertainty below 1\% in less than 3 minutes of proton irradiation at the given currents, with negligible environmental background. 

At a proton beam energy $E_p = 1747$~keV, the average target thickness was found to be 5~keV, corresponding to a physical thickness of about 170~nm and to an areal density $N_{^{13}\rm C}\approx10^{18}$ atoms/cm$^2$. 
The heights of the yield plateau of all fresh targets were consistent with each other within experimental uncertainties, indicating that all targets had the same initial stoichiometry and confirming the reproducibility of the evaporation procedure. 

For some targets, the thickness uniformity was also verified by repeating the resonance scan on three different spots of the same target, 6~mm apart from each other. 
This requirement was especially important for the LUNA experiment because the $\alpha$-particle beam has a typical diameter of about 15~mm on target, so uniformity of the evaporated layer had to be guaranteed over the whole beam-spot area. In the three spots examined, the shapes of the resonance profile were consistent within the uncertainties \cite{Ciani2017}.
Based on the test measurements, no modification of stoichiometry was observed during irradiation at Atomki.
In addition, NRRA was performed also on a few natural carbon targets, whose $^{13}$C content is known to be 1.1\%.
The comparison of the plateau heights confirmed a $^{13}$C  abundance in the enriched targets compatible with the 99\% value guaranteed by Sigma Aldrich \cite{CianiPhD}.

\subsection{The NRRA results}
\label{sec:NRRA-results}
\begin{figure}[t!]
\begin{center}
\resizebox{1\columnwidth}{!}{
\begin{tabular}{c c}
\includegraphics{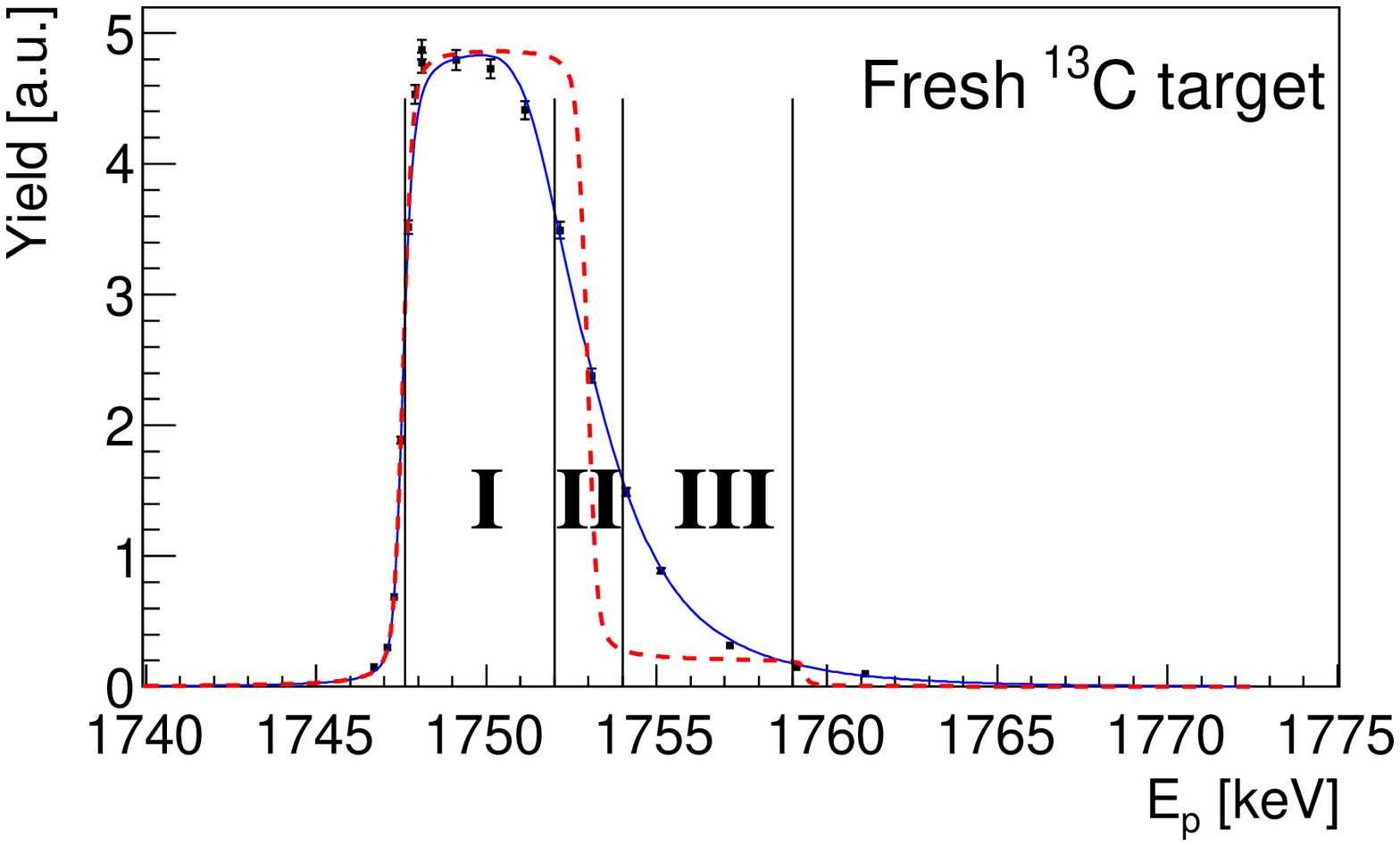}\\
\includegraphics{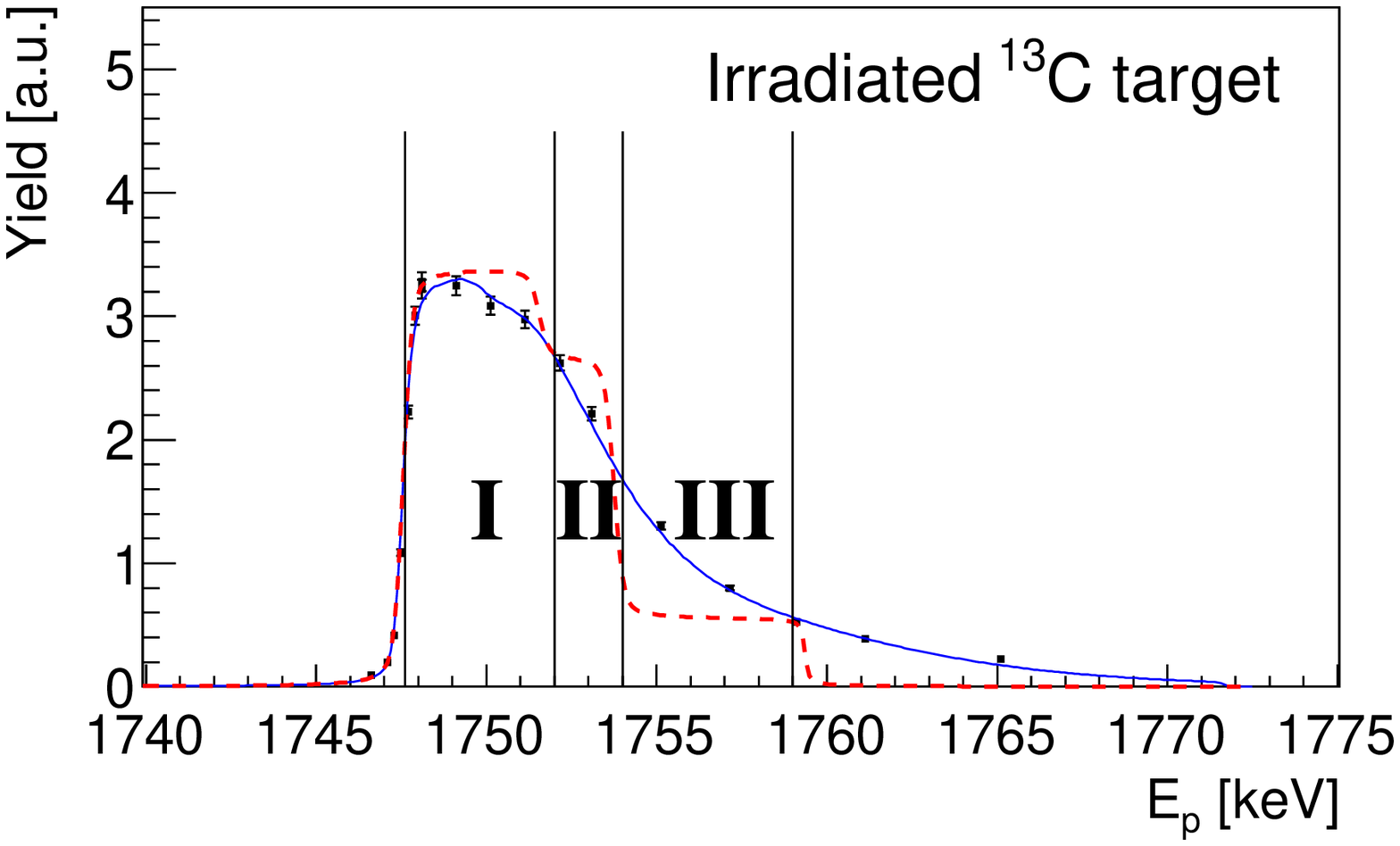}\\
\end{tabular}}
\end{center}
\caption{$^{13}$C(p,$\gamma$)$^{14}$N thick-target resonance yields obtained on a fresh $^{13}$C target (upper panel) and on the same target after 2.1 C of accumulated $\alpha$-beam charge (lower panel).
Experimental data (black squares) were fit taking into account 
beam spread with (blue line) and without (red dashed line) beam straggling effects. Vertical lines indicate the boundaries of  layers with different stoichiometries (see text for details).}
\label{fig:Yield_curve_NRRA_final}
\end{figure}

Figure \ref{fig:Yield_curve_NRRA_final} shows a typical resonance yield curve obtained on a fresh target (upper panel) and on a target exposed to about 2.1~C of $\alpha$-beam irradiation at LUNA (lower panel). As can be seen, the shapes of the resonance profiles differ significantly as a result of beam exposure, both in height and FWHM of the yield plateau.

In order to quantify the degree of deterioration, expe\-ri\-mental data (black points in fig. \ref{fig:Yield_curve_NRRA_final}) were fit taking into account a number of experimental effects, such as beam energy resolution and beam straggling \cite{StragglingBemmerer} within the target.
These factors can be folded into the expression of the yield (eq. \ref{eq:cross_sect}) as \cite{Iliadis}:

\begin{equation}
\label{NRA_yield}
\resizebox{1\columnwidth}{!}{
$Y(E_0) =k
\int_{E_0-\Delta E}^{E_0}{\rm d}E' 
\int_{E_i=0}^\infty {\rm d}E_i 
\int_{E=0}^{E_i} \frac{\sigma(E)}{\epsilon_{\rm eff}(E)}
g(E_0,E_i)
f(E_i,E,E')
{\rm d}E$}
\end{equation}

Here, $k$ is a normalization constant that includes the branching ratio of the transition and the $\gamma$-ray detection efficiency at the resonance energy; $g(E_0,E_{i})dE_i$ describes the energy distribution of particles in the beam; and $f(E_{i},E,E')dE$ describes the beam energy loss and straggling through the
target (see \cite{Iliadis} for more details).
Provided all other quantities are known, a measurement of the resonance yield profile can be used to determine $\epsilon_{\rm eff}(E)$ (\textit{i.e.}, the stoichiometric ratio $N_{\rm I}/N_{\rm A}$) at the resonance energy.

For the present analysis, the $^{13}$C(p,$\gamma$)$^{14}$N reaction cross-section $\sigma$(E) was taken from the TENDL-2017 nuclear data library \cite{TENDL2017} 
and evaluated as the sum of a non-resonant and a resonant component described by a second order polynomial and the Breit-Wigner formula, respectively. 
The stopping power $\epsilon_{\rm eff}(E)$ was assumed to be constant over the total width ($\Gamma \simeq 135$~eV) of the resonance. 
The $g(E_0,E_i){\rm d}E_i$ function was assumed to follow a normal distribution with a FWHM of $\sim 350$~eV \cite{Rajta2018}.

As for the calculation of the $f(E_i,E,E'){\rm d}E$ function, assumptions on some target properties were needed. 
Here, it was assumed that the targets initially consisted of $^{13}$C and Ta only, but with varying stoichiometric ratios as a function of depth. 
The TRIM software \cite{TRIM} was then used to calculate the energy loss and energy straggling of the beam for a given $N_{\rm Ta}/N_{^{13}\rm C}$ ratio. 
It was found that the resonance profiles could be well-reproduced by assuming three layers of different $N_{\rm Ta}/N_{^{13}\rm C}$ stoichiometric ratios (calculated using a $\chi^2$ minimization), but with homogeneous composition within each layer.
For fresh targets we assumed $N_{\rm Ta}/N_{^{13}\rm C}=0$.
The calculated yield curves obtained including the beam spread with and without beam straggling effects are shown in fig. \ref{fig:Yield_curve_NRRA_final} as solid (blue) and dashed (red) lines, respectively. 
Vertical lines indicate the boundaries of the layers with different stoichiometries.
In fitting the yield profiles, the 
$N_{\rm Ta}/N_{^{13}\rm C}$ ratios in the various layers and their thickness were treated as free parameters. 

\begin{figure}[t!]
\begin{center}
\resizebox{1\columnwidth}{!}{
\includegraphics{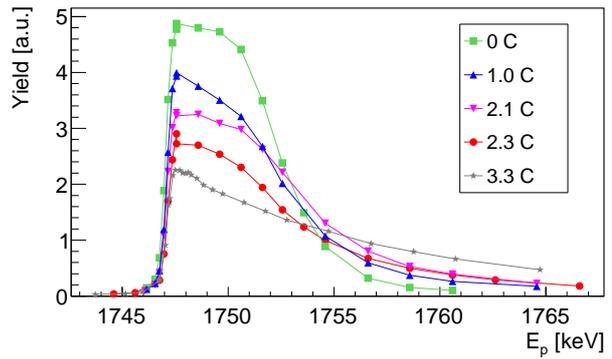}}
\end{center}
\caption{(Colour online) Resonance yield profiles measured on targets with different accumulated ($\alpha$-beam) charge. Lines are drawn to guide the eye.}
\label{fig:NRRA_profile3}
\end{figure}

Finally, NRRA measurements were also repeated on a sample of four targets after target irradiation at LUNA with different accumulated charges. The results from these measurements were used to validate the $\gamma$-shape analysis method (see sect. \ref{sec:validation}).
Figure \ref{fig:NRRA_profile3} shows the NRRA profiles obtained on targets with different amounts of accumulated $\alpha$-beam charges. 
A significant modification of the resonance yield curve was observed with increasing accumulated charge during $\alpha$-beam irradiation at LUNA.
In particular, the plateau becomes lower and the falling edge becomes longer. 
The observed depth profile indicates strong diffusion of $^{13}$C into the Ta backings. However, the position of the leading edge of the yield curves does not change appreciably, indicating negligible carbon build-up on target surface during irradiation.

The extracted effective stopping power values given in table \ref{tab:table1} (see sect. \ref{sec:errors} for an evaluation of the uncertainties) are those corresponding to layer I\footnote{Indeed, as the cross-section of $^{13}$C($\alpha$,n)$^{16}$O reaction drops exponentially with energy, the outermost layers of the target gives the main contribution to the reaction yield. Thus, only the stoichiometric ratio of layer I are of interest here.}.

\begin{table}[t!]
\caption{Stoichiometric ratios fitted with NRRA (third column) and corresponding effective stopping power values [keV/10$^{18}$ atoms~cm$^{-2}$] on targets with different accumulated ($\alpha$-beam) charge.}
\begin{center}
\resizebox{1\columnwidth}{!}{
\begin{tabular}{|c | c | c | c | }
\hline
Target	&	Charge [C] & $N_{\rm Ta}/N_{\rm ^{13}C}$ & $\epsilon_{\rm eff}\pm \Delta \epsilon_{\rm eff}$ \\ \hline
T29	 	&0	&	0.000 &3.12$\pm$0.16 \\
T26		&1.00 & 0.047$\pm$0.001 &3.84$\pm$0.19 \\
T29 	&2.10 & 0.101$\pm$0.002 &4.62$\pm$0.23 \\
T28		&2.34 &	0.149$\pm$0.003 &5.41$\pm$0.27 \\
MT10    &3.30 &	0.202$\pm$0.004 &6.32$\pm$0.32 \\ \hline
\end{tabular}
\label{tab:table1}
}
\end{center}
\end{table}

\section{The $\gamma$-shape analysis method}
\label{sec:methodGamma}
For a non-resonant radiative capture reaction A$(x,\gamma)$B at sub-Coulomb energies, the shape of a primary $\gamma$-ray transition is governed by the behaviour of the reaction cross-section
$\sigma(E)$ over the energy range covered by the incident beam as it loses energy in traversing the target \cite{Imbriani2005}.
For a thick target, the shape is also influenced by the energy dependence of the stopping power, and by the concentration profile of active nuclei as a function of target depth (which may change during irradiation).

Additional experimental effects may further contribute to the exact shape of the 
$\gamma$-ray line and must be taken into account.
Specifically, the high-energy rise of the peak may be Doppler-shifted by the recoil of the compound nucleus, while its low-energy tail may be affected by beam straggling effects.
Thus d$Y_i$, the number of counts per unit of charge in channel $i$ of the acquired $\gamma$-ray spectrum, with central value $E_{\gamma_i}$ (E$_{i}$ is the corresponding projectile energy) and width $\Delta E_\gamma$ is given by the expression \cite{Imbriani2005}:
\begin{equation}
\label{eq:Yield}
{\rm d}Y_i = A \frac{\sigma(E_{i})}{\epsilon_{\rm eff}(E_{i})}\Delta E_\gamma \zeta(E_\gamma) P(E_{i}) f(E_i,E,E') dE_i, 
\end{equation}
where $A$ is a normalization constant that includes the branching ratio of the transition and the $\gamma$-ray detection efficiency, $\zeta(E_\gamma)$ is a Gaussian function accounting for the energy resolution of the $\gamma$-ray detector, 
$P(E_{i})$ describes the concentration profile of active nuclei within the target (see below), and $f(E_i,E,E'){\rm d}E_i$ describes the energy broadening due to beam straggling effects.

The target concentration profile $P(E)$ can be modelled as the product of two Fermi functions \cite{Caciolli}: 
\begin{equation}
\label{eq:Fermi}
P(E)=\Big[\textrm{exp}\Big(\frac{E-E_0}{\delta_1}\Big)+1\Big]^{-1}\Big[\textrm{exp}\Big(\frac{E_0-E-\Delta E}{\delta_2}\Big)+1\Big]^{-1}
\end{equation}
where $E_0$ is the incident beam energy, $\Delta E$ the target thickness, and $\delta_1$ and $\delta_2$ are two parameters accounting, respectively, for the slopes of the falling and leading edges of the thick-target profile. 

The analysis of $\gamma$-ray line shapes has been extensively used in the past to extract information on unknown cross-sections of astrophysical reactions (\cite{15Npg,Scott,DiLeva}), provided that the target profile $P(E)$ could be measured independently (\textit{e.g.}, through NRRA analysis). 

In the present study, we exploited instead the $\gamma$-shape analysis approach to determine $P(E)$ and the effective stopping power $\epsilon_{\rm eff}$ using the well-known cross-section of the $^{13}$C(p,$\gamma$)$^{14}$N reaction, as explained in the following sections.

\subsection{The $\gamma$-shape measurements at LUNA}
\label{sec:gamma-LUNA}

In order to monitor the target degradation during the $^{13}$C($\alpha$,n)$^{16}$O measurements, data taking at LUNA  consisted of long $\alpha$-beam runs with accumulated charges of $\approx 1$~C per run, interspersed by short proton-beam runs with typical accumulated charges of 0.2~C at most, so as to minimize possible changes in target stoichiometry caused by the proton irradiation itself.

Proton beam runs were all performed at the same reference energy, $E_p=310$~keV. The choice for this energy was dictated by the need to maximize counting statistics while minimizing beam-induced background from a broad resonance at $E_p \simeq 340$~keV in the $^{19}$F(p,$\alpha \gamma$)$^{16}$O reaction on ever present $^{19}$F contaminants in the experimental setup. Note that at such a low proton-beam energy the resulting target thickness is $\Delta E \simeq 15$~keV 
and neither the $^{13}$C(p,$\gamma$)$^{14}$N reaction cross-section nor the effective stopping power $\epsilon_{\rm eff}$ can be regarded as constant.

Primary $\gamma$ rays ($E_\gamma = 7840~$keV) arising from the $^{13}$C(p,$\gamma$)$^{14}$N direct capture transition into the $^{14}$N ground state (hereafter, DC~$\rightarrow$~GS transition) were detected using a HPGe detector with a relative efficiency of 120\% and FWHM of 2.8 keV at $E_\gamma=1460$~keV. 
The detector was mounted at 55\textdegree\ to the beam axis and brought to a distance of 5 mm from the target holder \cite{Csedreki2017}.
The same type of electronics and DAQ used in the NRRA measurement was used to acquire the $\gamma$-spectrum at LUNA.

\subsection{The $\gamma$-shape analysis and results}
\label{sec:gamma-results}

\begin{figure}[t!]
\begin{center}
\resizebox{.85\columnwidth}{!}{
    \begin{tabular}{c c}
    \includegraphics{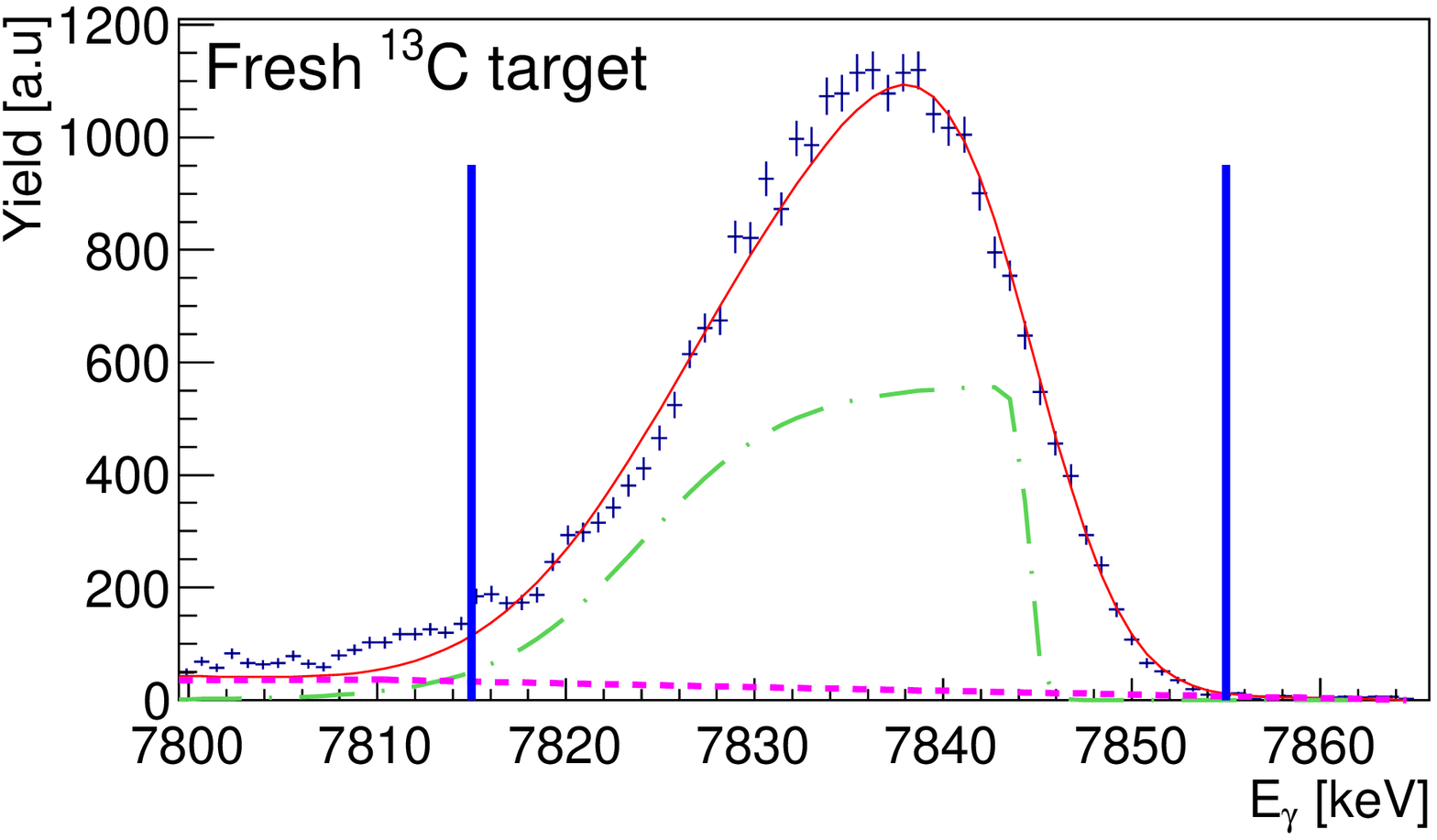}\\
    \includegraphics{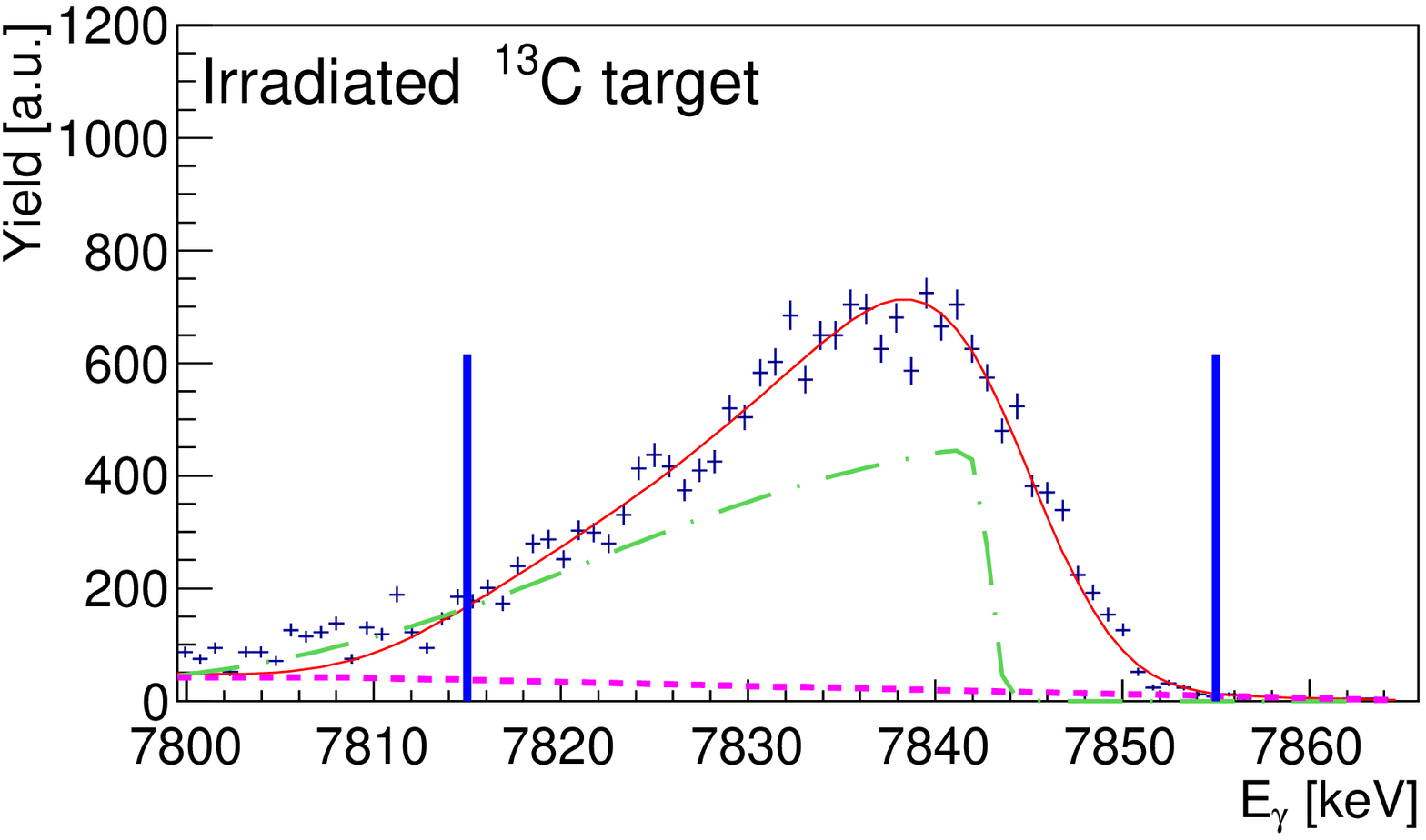}\\
    \end{tabular}}
\end{center}
\caption{(Colour online) Gamma-ray peak from the  $^{13}$C(p,$\gamma$)$^{14}$N DC$\rightarrow$GS transition as  obtained on a fresh $^{13}$C target (upper panel) and on the same target irradiated with $\alpha$-beam for 3.3~C of accumulated charge (lower panel).
Experimental data were fit (red line) using eq. \ref{eq:Yield} and including a linear background (dashed line). The $\chi^2$ was minimized in the region delimited by the vertical lines.
In dash-dotted green, the target profile $P(E)$ (in arbitrary units), as defined in eq. \ref{eq:Fermi}, shows the concentration of active nuclei as a function of depth (\textit{i.e.} beam energy within the target).}
\label{fig:fit}
\end{figure}

Figure \ref{fig:fit} shows the ${\rm DC} \rightarrow {\rm GS}$ peak ($E_\gamma = 7840~$keV) of two $\gamma$-ray spectra acquired on a fresh target (upper panel) and after an $\alpha$-beam irradiation of 3.3 C of total accumulated charge (lower panel).

Experimental spectra (blue crosses) were fit using eq. (\ref{eq:Yield}), where the low-energy trend of the $^{13}$C(p,$\gamma$)$^{14}$N reaction cross-section was taken from King \textit{et al.} \cite{KING1994354} and Genard \textit{et al.} \cite{13Cpg14N_res511}, and the beam straggling distribution function $f(E_i,E,E'){\rm d}E_i$ was evaluated by Monte Carlo simulations using TRIM.

For runs on fresh targets, parameters $A$, $\Delta E$, $\delta_1$ and $\delta_2$ in eq. \ref{eq:Yield} and \ref{eq:Fermi} were left free to vary, while the stoichiometric ratio $N_{\rm Ta}/N_{\rm ^{13}C}$ was set to 0, as no degradation had yet occurred.
For runs on irradiated targets, parameters $A$ and $\delta_1$ were fixed to the fit values of the ``fresh'' target, leaving $\Delta E$, $\delta_2$ and $N_{\rm Ta}/N_{\rm ^{13}C}$ as free parameters.

The results of the fitting procedure are shown as red curves in Fig. \ref{fig:fit}, while dash-dotted green curves show target profiles $P(E)$ defined in eq. (\ref{eq:Fermi}) in arbitrary units. Note the change in the shape of the target profile $P(E)$ following irradiation with the $\alpha$-beam.
A linear background (dashed line) was included in the ROI of the fit to account for multiple Compton-scatter events in the HPGe detector.
The $\chi^2$ was minimized in the region delimited by vertical lines, for a number of degrees of freedom $\nu=40$. We obtained a reduced $\tilde{\chi}^2 \approx 1.6$ for both plots shown.

Table \ref{tab:parameters} reports the values of the fit parameters for both spectra shown in fig. \ref{fig:fit}. As expected, $N_{\rm Ta}/N_{\rm ^{13}C}$ and $\delta_2$ show a significant change, indicating a strong modification in the target stoichiometry and a likely diffusion of $^{13}$C nuclei into the backing.

\begin{table}[h!]
\caption{Parameter values obtained from the $\gamma$-shape fits to the peaks in fig. \ref{fig:fit}. The normalization constant is $A = (3.16\pm0.01) \times 10^{-4}$ (in a.u.) for both profiles.}
\label{tab:parameters}
\resizebox{1\columnwidth}{!}
{\begin{tabular}{|c|c|c|c|c|}
\hline
$Q$ [C] & $N_{\rm Ta}/N_{\rm ^{13}C}$  & $\Delta E$ [keV] & $\delta_1$ [keV]   & $\delta_2$  [keV]   \\ \hline
0             & 0       & 21.7$\pm$0.1  & 0.33$\pm$0.03   & 4.13$\pm$0.22  \\ \hline
3.3           & 0.16$\pm$0.011  & 22.2$\pm$0.1  & 0.33$\pm$0.03   & 10.19$\pm$0.22 \\ \hline
\end{tabular}}
\end{table}

\section{Validation and discussion}
\label{sec:validation}
\subsection{Role of inactive nuclides in the $\gamma$-shape analysis}
\label{sec:independency}
In order to check the effect of possible light contaminants (\textit{e.g.}, H, He, C, O) on the effective stopping power, we performed several SRIM calculations for proton energies $E_p = 280 - 310~$keV, and alpha energies $E_{\alpha} = 300-400~$keV 
(relevant to the $^{13}$C($\alpha$,n)$^{16}$O data taking campaign). 
In the energy ranges considered, the energy dependence of stopping power for each element (H, He, C, O), assumed as the only contaminant in the target, changes by less than 3\% for proton projectiles and less than 5\% for alpha particles. Similar conclusions can be drawn in the case where more than one contaminant is present at the same time.
These conclusions were further supported by additional ERDA analysis performed on irradiated targets at the Ion Beam Center of Helmholtz-Zentrum Dresden-Rossendorf. The analysis confirmed that the concentration of elements such as H, He and O after the $\alpha$-beam irradiation at LUNA was at most 10\%
 \cite{ERDA}.
We conclude that, for our $\gamma$-shape analysis, the effective stopping power is essentially insensitive to the actual species of inactive nuclei present in the target \cite{CianiPhD}.
Stoichiometric values $N_I/N_A$ obtained from the $\gamma$-shape fit are reported in table \ref{tab:diffcont} for each one of the inactive species considered, together with the associated stopping powers for proton and $\alpha$ beams. 

\begin{table}[h!]
\caption{Stoichiometric ratios for possible inactive nuclei (H, He, C, O and Ta), as obtained from a $\gamma$-shape fit of the primary $\gamma$ ray in $^{13}$C(p,$\gamma$)$^{14}$N. The corresponding effective stopping powers are calculated for a proton beam at $E_{\rm p} = 310$~keV and an $\alpha$ beam at $E_{\rm \alpha} = 400$~keV.}
\resizebox{1\columnwidth}{!}{
\begin{tabular}{|c|c|c|c|c|}
\hline
Inactive  & $N_I/N_A$     & $\epsilon_{\rm eff} (p)$ &  $\epsilon_{\rm eff} (\alpha)$\\ 
species & ~& [{\rm keV}/$10^{18}${\rm atoms/cm}$^2$] & [{\rm keV}/$10^{18}${\rm atoms/cm}$^2$] \\\hline
H                & 1.92$\pm$0.15 & 14.36$\pm$1.44  & 59.96$\pm$3.21                    \\ \hline
He                & 1.19$\pm$0.11 & 14.51$\pm$1.45  & 54.79$\pm$4.14                    \\ \hline
$^{12}$C                & 0.55$\pm$0.023 & 14.51$\pm$1.45  & 57.67$\pm$3.15                    \\ \hline
O                & 0.48$\pm$0.016 & 14.65$\pm$1.46 &  56.65$\pm$2.80                    \\ \hline
Ta               & 0.16$\pm$0.011 & 14.53$\pm$1.45 & 53.45$\pm$2.64                    \\ \hline
\end{tabular}}
\label{tab:diffcont}
\end{table}

\subsection{Comparison of NRRA and $\gamma$-shape analysis results}
To validate the results of the $\gamma$-shape analysis approach, a comparison to the results obtained with the well-established NRRA method was made.
To this end, the effective stopping powers arising from the stoichiometric ratios obtained with the NRRA at $E_p= 1747.6$~keV (table \ref{tab:table1}) were recalculated at $E_p =310$~keV using eq. (\ref{eq:eff_stop}) assuming that the targets consist of a compound of only $^{13}$C and Ta.

Table \ref{tab:Exp_data_eff_stop} reports the values of the effective stopping powers obtained with the two methods for different accumulated ($\alpha$-beam) charges. The results obtained are in agreement within uncertainties (see sect. \ref{sec:errors} for the uncertainties evaluation).
\begin{table}[h!]
\caption{Effective stopping powers [$\rm{keV}/10^{18}\ \rm{atoms}/\rm{cm}^{2}$] calculated at $E_{\rm p} = 310$~keV using the NRRA and the $\gamma$-shape analysis approach for targets of different accumulated ($\alpha$-beam) charge.}
 \begin{center}
\resizebox{0.85\columnwidth}{!}{
\begin{tabular}{|c|c|c|c|}
\hline
Target	&	Charge	& $\gamma$-shape		& NRRA  \\ 
&			[C]	& $\epsilon_{\rm eff}\pm \Delta \epsilon_{\rm eff}$ 	& $\epsilon_{\rm eff}\pm \Delta \epsilon_{\rm eff}$  				\\ \hline
T29		&0	&9.38$\pm$0.48&	9.37 $\pm$0.47 \\
T26		&1.00	&10.53$\pm$1.05& 10.83$\pm$0.54 \\
T29 		&2.10&12.15$\pm$1.21&12.51$\pm$0.63 \\
T28		&2.34	&13.49$\pm$1.35&	14.01$\pm$0.70 \\
MT10	&3.30	&14.53$\pm$1.45&	15.64$\pm$0.78 \\ \hline
\end{tabular}}
\end{center}
\label{tab:Exp_data_eff_stop}
\end{table}

\subsection{Uncertainties budget}
\label{sec:errors}
The overall uncertainty on the effective stopping power evaluation has three main contributions: a 4-5\% systematic error on SRIM tabulated values of stopping powers for pure materials (common to both methods); a 1\% and a 3\% systematic error on the charge integration on target for measurements at Atomki (NRRA) and LUNA ($\gamma$-shape), respectively; a 2\% and an 8\% fit uncertainty on the extracted stoichiometric ratios from the NRRA and the $\gamma$-shape approaches, respectively. In both approaches, fit uncertainties were calculated \cite{PhysRevD.98.030001} by varying the $N_{\rm Ta}/N_{\rm ^{13}C}$ within a range $[N_{\rm Ta}/N_{\rm ^{13}C} \pm \delta]$ until the $\chi^2$ value increased by a fixed amount $\Delta \chi^2$ (which depends on the number of fit parameters, 3.2 in this specific case) around its minimum value.
The overall uncertainty on the effective stopping power was then obtained by summing in quadrature all sources of errors and resulted in an overall 5\% error for the NRRA measurements and an overall 10\% error for the $\gamma$-shape analysis, respectively. 
A summary of the main uncertainties for the two techniques is presented in table \ref{tab:uncertainties}.
\begin{table}[htb]
\caption{Summary of main uncertainties for the two techniques.}
\resizebox{1\columnwidth}{!}{
\begin{tabular}{|c|c|c|}
\hline
Uncertainty source & NRRA & $\gamma$-shape \\ \hline
Charge accumulation & 1\%  & 3\% \\ \hline
Stopping power from SRIM & 5\%  & 5\% \\ \hline
Evaluation of stioichimetric ratio & 2\%  & 8\% \\
\hline
Total uncertainty & 5\% & 10\% 
\\ \hline
\end{tabular}}
\label{tab:uncertainties}
\end{table}

\section{Target degradation correction applied to the evaluation of the $^{13}$C($\alpha$,n)$^{16}$O reaction cross-section}
\label{sec:cross-section}
During the $^{13}$C($\alpha$,n)$^{16}$O data taking campaign, over a hundred $^{13}$C targets were used for an overall accumulated $\alpha$-beam charge of about 300~C. For each target, $\gamma$-ray spectra acquired during short proton runs, taken before and after long $\alpha$-beam irradiation runs, were analyzed following the procedure described in sect. \ref{sec:methodGamma}  to  correct for target degradation effects in the evaluation of the $^{13}$C($\alpha$,n)$^{16}$O reaction cross-section.

\begin{figure}[t!]
\begin{center}
\resizebox{1\columnwidth}{!}{
\includegraphics{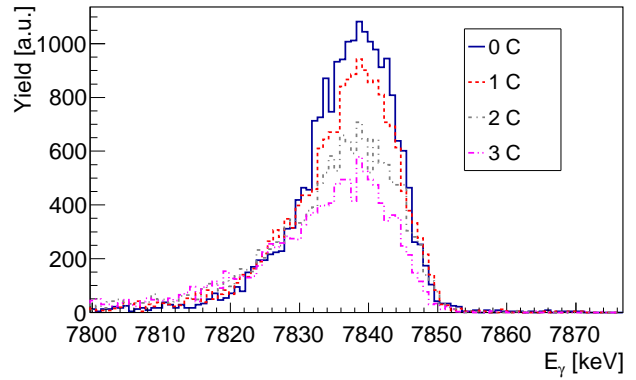}}
\end{center}
\caption{(Colour online) Overlay of $\gamma$-ray spectra for the $^{13}$C(p,$\gamma$)$^{14}$N DC~$\rightarrow$~GS transition acquired on the same target at different accumulated ($\alpha$-beam) charge. Both the height of the peak and its FWHM change with increased charge as expected, indicating severe target modification during $\alpha$-beam irradiation.}
\label{fig:Cpg_comparison}
\end{figure}

As an example, fig. \ref{fig:Cpg_comparison} shows a superposition of the DC~$\rightarrow$~GS peak in four $\gamma$-ray spectra acquired on the same target at increasing values of accumulated ($\alpha$-beam) charge. As expected, the higher the accumulated charge, the lower the (p,$\gamma$) yield and the broader the shape of the $\gamma$-ray line.

\begin{figure}[t!]
\begin{center}
\resizebox{1\columnwidth}{!}{
\includegraphics{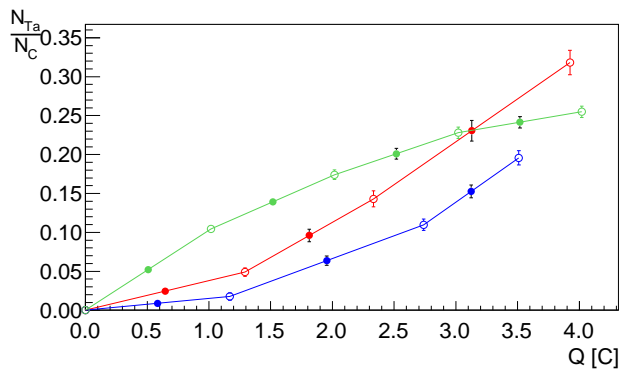}}
\end{center}
\caption{(Colour online) Stoichiometric ratios $N_{\rm Ta}/N_{^{13}\rm C}$ as a function of accumulated charge $Q$ on three different targets (represented by different colours). Open symbols correspond to values obtained with the $\gamma$-shape analysis; filled symbols represent linearly interpolated values.}
\label{fig:degrad_vs_Q}
\end{figure}

From fits to each peak, we extracted values of $N_{\rm Ta}/N_{\rm ^{13}C}$ at the target surface and plotted them as a function of the accumulated charge $Q$ (see fig. \ref{fig:degrad_vs_Q}, where for clarity, results are displayed for three targets only). Open symbols in the figure correspond to stoichiometric ratios determined with the $\gamma$-shape analysis method on reference proton runs, while filled symbols correspond to linearly interpolated values. The latter were used to calculate average effective stopping powers (eq. 3) to be used in the evaluation of the $^{13}$C($\alpha$,n)$^{16}$O reaction cross-section, thus accounting for target degradation during long $\alpha$-beam irradiation in-between successive proton runs.

\begin{figure}[t!]
\begin{center}
\resizebox{.99\columnwidth}{!}{
\includegraphics{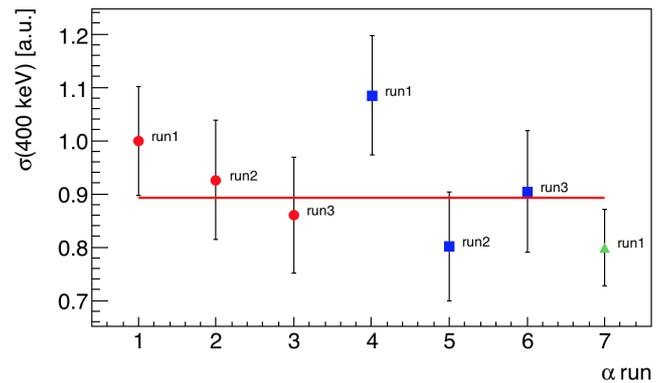}}
\end{center}
\caption{(Colour online) $^{13}$C($\alpha$,n)$^{16}$O cross-section (in a.u.) extracted from different $\alpha$-beam runs on three different targets (indicated in red, blue and green). The solid red line represents the weighted average of the data points shown. Error bars include statistical and systematic uncertainties in the $\epsilon_{\rm eff}$ evaluation.}
\label{fig:CS}
\end{figure}

Figure \ref{fig:CS} shows the  $^{13}$C($\alpha$,n)$^{16}$O cross-sections (in arbitrary units) evaluated from measurements performed at the same beam energy ($E_{\alpha}=400$~keV) on three different targets. 
The error bars shown arise from a combination of statistical and systematic uncertainties in the $\epsilon_{\rm eff}$ evaluation. All data points are within 2$\sigma$ from the weighed average (red line). 
Final results on the $^{13}$C($\alpha$,n)$^{16}$O cross-section over the full energy range ($E_{\alpha} =305-400$~keV) covered at LUNA will be presented in a forthcoming publication.

\section{Conclusions}
In this paper we reported on a new application of the $\gamma$-shape analysis used to monitor {\em in situ} the degradation of $^{13}$C targets exposed to intense $\alpha$-beam irradiation during the $^{13}$C($\alpha$,n)$^{16}$O reaction study at LUNA.
Specifically, fits to the peak shape of the DC~$\rightarrow$~GS $\gamma$-ray transition in the $^{13}$C(p,$\gamma$)$^{14}$N reaction were used to obtain quantitative information on target degradation as a function of accumulated ($\alpha$-beam) charge on target.

The $\gamma$-shape analysis was used as an alternative to the standard NRRA, whose application at LUNA was precluded by the lack of appropriate resonances in the energy range accessible with the 400~kV accelerator.
NRRA measurements were, instead, performed at Atomki, both to characterize initial target thickness and stoichiometry and, for a subset of targets, as a way to validate the $\gamma$-shape analysis.
A comparison of the stoichiometric values obtained with both methods shows agreement within experimental uncertainties. 

We also verified that the effective stopping powers used in the evaluation of the $^{13}$C($\alpha$,n)$^{16}$O reaction cross-sections were independent from the assumption of inactive contaminant(s) present in the target.

The effective stopping power values obtained with the $\gamma$-shape analysis were extracted with an overall 10\% uncertainty.
While the use of the $\gamma$-shape analysis was validated specifically for $^{13}$C targets in the present study, this approach may have wider applications especially where the use of traditional analytical methods, such as the NRRA is not possible.

\section*{Acknowledgement}
The authors would like to thank Donatello Ciccotti and the LNGS and INFN Naples mechanical workshops for technical support.
The authors are grateful to Jaakko Julin and Frans Munnik for ERDA measurements at the Ion Beam Center of Helmholtz-Zentrum Dresden Rossen\-dorf (HZDR).
Support from the National Research, Development and Innovation Office
NKFIH (contract numbers PD 129060 and K120666) is also acknowledged, as well as funding from STFC (ST/P004008/1), DFG (BE 4100/4-1), HGF (ERC-RA-0016) and the grant STAR from the University of Naples/Compagnia di San Paolo. This work is supported by “ChETEC” COST Action (CA16117).
This is a pre-print of an article published in European Physical Journal A. The final authenticated version is available online at: https://doi.org/10.1140/epja/s10050-020-00077-0.
%
 \bibliographystyle{epj.bst}
 \bibliography{mainArxiv}
%

\end{document}